\documentclass{article}

\usepackage{microtype}
\usepackage{graphicx}
\usepackage{subfigure}
\usepackage{booktabs} 

\usepackage{amsmath}
\usepackage{amssymb}
\usepackage{mathtools}
\usepackage{amsthm}

\usepackage{hyperref}
\usepackage{float}
\usepackage{amsfonts}

\newcommand{\mE}{\mathbb{E}}
\newcommand{\bx}{\mathbf{x}}
\newcommand{\by}{\mathbf{y}}
\newcommand{\bz}{\mathbf{z}}

\usepackage[accepted]{icml2024}

\usepackage[capitalize,noabbrev]{cleveref}

\theoremstyle{plain}

\theoremstyle{definition}

\theoremstyle{remark}

\usepackage[textsize=tiny]{todonotes}

\icmltitlerunning{ Integrating Latent Variable and Auto-regressive Models for Enhanced Goal Directed Generation}

\begin{document}

\twocolumn[
\icmltitle{Rethinking Molecular Design: Integrating Latent Variable and Auto-Regressive Models for Goal Directed  Generation}



\begin{icmlauthorlist}
\icmlauthor{Heath Arthur-Louis}{sch}
\icmlauthor{Amina Mollaysa}{yyy,comp}
\icmlauthor{Michael Krauthammer}{yyy,comp}
\end{icmlauthorlist}

\icmlaffiliation{yyy}{Department of Quantitative Biomedicine, University of Zurich, Switzerland}
\icmlaffiliation{comp}{ETH AI Center}
\icmlaffiliation{sch}{ETH, Zurich, Switzerland}

\icmlcorrespondingauthor{Heath  Arthur-Louis}{contact@louis-heath.me}
\icmlcorrespondingauthor{Amina Mollaysa}{maolaaisha.aminanmu@uzh.ch}

\icmlkeywords{ Molecule Generation, Drug Design, Conditional Generation, De Novo, VAE}

\vskip 0.3in
]



\printAffiliationsAndNotice{}  

\begin{abstract}
De novo molecule design has become a highly active research area, advanced significantly through the use of state-of-the-art generative models. Despite these advances, several fundamental questions remain unanswered as the field increasingly focuses on more complex generative models and sophisticated molecular representations as an answer to the challenges of drug design.  In this paper, we return to the simplest representation of molecules, and investigate overlooked limitations of classical generative approaches, particularly Variational Autoencoders (VAEs) and auto-regressive models. We propose a hybrid model in the form of a novel regularizer that leverages the strengths of both to improve validity, conditional generation, and style transfer of molecular sequences. Additionally, we provide an in depth discussion of overlooked assumptions of these models' behaviour. 

\end{abstract}





\section{Introduction}
Molecule discovery tasks can be divided into two types. Global optimization aims to find molecules with a specific target property. Local optimization starts with an initial molecule and searches for similar molecules that have the desired property without deviating too much from the original. Past advances in molecular design have primarily leveraged either latent variable models or auto-regressive models, each with distinct advantages and shortcomings. While auto-regressive models excel in capturing conditional distributions and generating valid molecules with desired properties, they tend to over-fit the training data, leading to a limited number of novel molecules when trained on small datasets. Additionally, they lack the flexibility of latent variable models, which facilitate local optimization and complex transformations such as style transfer. On the other hand, latent variable models often struggle to generate valid molecules unless more restrictive and complex representations are applied, due to the challenges of generating valid molecules from a compressed latent space \cite{DBLP:journals/corr/Gomez-Bombarelli16, kusner2017grammar, dai2018syntax, jin2018junction,mollaysa2019disentanagled}.

In this work, instead of relying on increasingly complex generative models and sophisticated representations as solutions for every challenge, we return to the simple string based representation; SMILES. By combining the strength of auto-regressive models and latent variable models, we aim to resolve the performance gap in conditional generation and enable effective style transfer without sacrificing the generative model's validity. 
We introduce a theoretical framework that combines the robust conditional distribution modeling capability of auto-regressive models with the flexible representational power of latent variable models. Our approach involves a dual-training mechanism where the auto-regressive model informs the training of the latent variable model, ensuring that the generative process respects both the desired property of the generated samples as captured by the auto-regressive model and the local structural coherence enforced by the latent variable.

The core of our methodology is a hybrid generative model that employs a conditional VAE architecture with an embedded auto-regressive model to guide the decoder (generative distribution). The model is trained using a novel objective function that incorporates a regularizer to ensure the generation of valid molecules with desired properties. The regularizer is implemented in two forms: a calibration regularizer, which is a Kullback-Leibler divergence term between the marginalized generative distribution and a target distribution defined by the auto-regressive model; and a reward-based regularizer, which rewards the generative model for producing molecules with high probability under the target distribution. Each regularizer guides the generative model, enhancing its ability to produce valid and conditionally appropriate molecules. 
Our contribution includes:
\begin{itemize}
\vspace{-0.5em}
    \item We present a comprehensive analysis of the limitations associated with latent generative models in molecular design, particularly focusing on their ability to perform conditional generation and style transfer.
    \vspace{-0.5em}
\item We propose a hybrid generative framework that effectively combines the strengths of auto-regressive and latent variable models, offering a balanced approach to molecular design.
\end{itemize}

\section{Methods}
\label{sec:method}
Suppose we are given a training set of pairs $\mathcal{D} = \{(\bx_i, \by_i)\}; i =1,\dots, N$,
where $\bx_i$ corresponds to a molecule and 
$\by_i$ represents its corresponding molecular properties. Goal directed molecule design can be formulated as learning the conditional distribution $p(\bx|\by)$ which allow us to set a value for $\by$ and generate a diverse set of molecules exhibiting the specified properties. Additionally, this task can be framed as local optimization, where we start with a prototype molecule and modify it to achieve the intended property without deviating (structurally) too far from the original molecule. We refer to this as style transfer over molecules, where we aim to modify 
$\bx$ to $\bx'$ such that $\bx'$ exhibits  property $\by'$. This can be achieved with a latent variable model 
 $p(\bx|\by,\bz)$ that allows us to control latent factors corresponding to certain molecular structures. The style transfer task can then be defined as follows:
\begin{align}
p_{\theta}(\bx' | \by', \bx) = \int p_\theta(\bx' | \by', \bz)q_{\phi}(\mathbf{z}|\mathbf{x}) \mathrm{d}\bz.
\end{align}
Concretely, this involves fitting a joint generative model of the form $p_{\theta}(\bx , \by, \bz)=   p_{\theta} (\bx |\by, \bz)p(\by)p(\bz)$
 and  a variational distribution $q_\phi(\bz | \bx)$ to infer the latent variable $\bz$. 
This model can be trained with the standard ELBO objective function:
\begin{align}
\label{elbo}
\mathcal{L}_{ELBO}(\theta, \phi) =
\sum_{i=1}^N \bigg\{&\mathbb{E}_{q_\phi(\bz_i | \bx_i)}[\log p_{\theta}(\bx_i |\by_i, \bz_i)] \nonumber \\
&- D_{KL}(q_\phi(\bz_i | \bx_i) || p(\bz_i)) \bigg\}
\end{align}
which corresponds to learning a supervised (conditional) VAE  \citep{kingma2014semi}.

Such models are often trained to generate molecules in one shot \cite{kusner2017grammar,dai2018syntax}, meaning the decoder assumes $p(\bx|\by,\bz) =\prod_{t=0}^T{p(x_t|\by,\bz)}$. This was set to avoid using a strong decoder (auto-regressive network trained with teacher forcing) that assumes $p(\bx|\by,\bz) =\prod_{t=0}^T{p(x_t|\by,\bz,  \mathbf{x_{<t}})}$ as it leads to the generative model \( p(\mathbf{x}|\mathbf{z}, \mathbf{y}) \) ignoring the latent variable \( \mathbf{z} \) and defaulting to \( p(\mathbf{x}|\mathbf{y}) \) \cite{DBLP:journals/corr/Gomez-Bombarelli16}. 
As a result, often the generative model  \( p(\mathbf{x}|\mathbf{z}, \mathbf{y}) \) is trained in a non-auto regressive fashion, which makes the task very challenging and often under performs compared to auto-regressive models that are designed to model  $p(\mathbf{x_t}|\mathbf{y}, \mathbf{x_{<t}})$  which is trained to predict the next token.

Additionally, the task of conditional generation becomes challenging when using standard conditional VAEs to model the generative distribution $p(\mathbf{x}|\mathbf{z}, \mathbf{y})$. The standard objective (Eq.\ref{elbo}) focuses only on reconstruction loss and a KL term, and has the potential to lead to scenarios where $p(\mathbf{x}|\mathbf{z}, \mathbf{y})$ may disregard $\mathbf{y}$, relying solely on $\mathbf{z}$ for molecule reconstruction. This often results in poor conditional generation performance, as the model effectively converges to $p(\mathbf{x}|\mathbf{z})$ and loses control over the properties of the generated molecules. Furthermore, the style transfer task also tends to perform poorly due to the lack of supervision during the training. The models are not exposed to examples where a latent 
$\bz$ from a prototype molecule combines with a new property 
$\by$  to predict what a molecule with structure 
$\bz$ and property $\by$ would look like. This gap in training results in the model's inability to effectively handle style transfer.

Our observations indicate that an auto-regressive model, which directly learns the conditional distribution \( p(\mathbf{x}|\mathbf{y}) \), demonstrates superior performance in conditional generation. This enhancement is attributed to its effective learning of the conditional distribution \( \tilde{p}(\mathbf{x}|\mathbf{y}) \approx p(\mathbf{x}|\mathbf{y}) \). By using this learned distribution as an approximation to the true conditional distribution, we can effectively guide our VAE model. This guidance enables the VAE to generate molecules that have a high probability under the learned distribution \( \tilde{p}(\mathbf{x}|\mathbf{y}) \).

\subsection{Proposed Regularizer}

We propose the use of a pre-trained  auto-regressive model $\tilde{p}(\mathbf{x}|\mathbf{y})$ as a surrogate for the true conditional distribution. This surrogate distribution can then be employed to direct the generative network, encouraging the generative model to generate molecules that have a high probability under the surrogate model. To achieve this, we augment the supervised VAE objective with one of the regularizer terms outlined below.
\vspace{-0.5em}
\paragraph{Calibration Regularizer} To achieve this constraint, we introduce a Kullback-Leibler (KL) divergence term that aligns the marginalized generative distribution \( p_{\theta}(\mathbf{x}|\mathbf{y}_i)  \) of a conditional VAE with the surrogate distribution \( \tilde{p}(\mathbf{x}|\mathbf{y}_i) \).
\begin{align}
\label{eq:sup_vae}
\mathcal{R}_1(\theta) =\min_{\theta}
D_{KL}(p_{\theta}(\mathbf{x}|\mathbf{y}_i)||\tilde{p}(\bx|\by_i) )
\end{align}
where the marginalized conditional distribution is given by:
\[
p_{\theta}(\mathbf{x}|\mathbf{y}_i) = \int p_{\theta}(\mathbf{x}|\mathbf{y}_i, \mathbf{z}) p(\mathbf{z}) \, d\mathbf{z}= \mathbb{E}_{\mathbf{z} \sim p(\mathbf{z})} p_{\theta}(\mathbf{x}|\mathbf{y}_i, \mathbf{z})
\]
Given we do not have direct access to the marginalized distribution \( p_{\theta}(\mathbf{x}|\mathbf{y}_i) \), we estimate it using the Monte Carlo integration approach: 
   $\hat{p}_{\theta}(\mathbf{x}|\mathbf{y}_i)\approx   \hat{p}_{\theta}(\mathbf{x}|\mathbf{y}_i) =\frac{1}{N} \sum_{n=1}^N p_{\theta}(\mathbf{x}|\mathbf{y}_i, \mathbf{z}_n)$
and then apply the Kullback-Leibler (KL) divergence on the approximated distribution:
\[
D_{KL}(\hat{p}_{\theta}(\mathbf{x}|\mathbf{y}_i)  \parallel \tilde{p}(\mathbf{x}|\mathbf{y}_i)) = \sum_{\mathbf{x}} \hat{p}_{\theta}(\mathbf{x}|\mathbf{y}_i) \log \frac{\hat{p}_{\theta}(\mathbf{x}|\mathbf{y}_i)}{\tilde{p}(\mathbf{x}|\mathbf{y}_i)}
\]
\vspace{-0.8em}
\paragraph{Reward-Based Regularizer}
As discussed in Section \ref{sec:experimental-results}, the calibration regualrizer encountered issues when paired with the standard character VAE decoder structure. To attempt to address this we formulated an alternative regularizer based on the following reward function:
\begin{align}
\label{eq:sup_vae}
\mathcal{R}_2(\theta) =\max_{\theta}
\sum_{i=1}^N \mE_{\bz\sim p(\bz)}\mE
_{\bx\sim p_{\theta}(\bx|\by_i,\bz)}\log \tilde{p}(\bx|\by_i)  \nonumber \\
=\max_{\theta}
\sum_{i=1}^N \sum_{j=1}^M\mE
_{\bx\sim p_{\theta}(\bx|\by_i,\bz_j)}\log \tilde{p}(\bx|\by_i)
\end{align}

This regularizer can be seen as  expected reward objective where $\log \tilde{p}(\bx|\by_i)$ can be seen as reward function and $p_{\theta}(\bx|\by_i,\bz_j)$ is the policy of the action, where we want to learn a policy model that maximizes our reward. 
The regularzier can be optimized using the  score function gradient estimate, i.e., $\mE_{\bx\sim p_{\theta}(\bx|\by_i,\bz_j)}\left[\log \tilde{p}(\bx|\by_i) \nabla_\theta \log p_{\theta}(\bx|\by_i,\bz_j)\right]$
However, the score function gradient estimator is known to be noisy and unstable unless we use large samples and control variate techniques. To avoid excessive computational complexity, we propose an alternative approach to achieve the same regularization \cite{mollaysa2020goal}. We reformulate our regularizer as \(\max_{\theta} \mathbb{E}_{\mathbf{x} \sim p_{\theta}(\mathbf{x}|\mathbf{y}_i,\mathbf{z}_j)} \tilde{p}(\mathbf{x}|\mathbf{y}_i)\), using \(\tilde{p}(\mathbf{x}|\mathbf{y}_i)\) as a positive reward function. This allows us to express the regularizer as:
\begin{align}
    \max_{\theta}&\mE
_{\bx\sim p_{\theta}(\bx|\by_i,\bz_j)} \tilde{p}(\bx|\by_i) \nonumber=\max_{\theta} \int  p_{\theta}(\bx|\by_i,\bz_j)\tilde{p}(\bx|\by_i) d\bx \nonumber \\
&=\max_{\theta}\mE_{\bx\sim \tilde{p}(\bx|\by_i)}p_{\theta}(\bx|\by_i,\bz_j)
\end{align}


\subsection{Incorporating the Regularizer}
We use the standard conditional VAE (Eq.\ref{elbo}) as our baseline and incorporated our regularizer to see if the regularizer would improve the generation, conditional generation as well as style transfer performance of the baseline model. 
We first train an LSTM-based model to learn $\hat{p}(\bx|\by)$. This pre-trained model is then served as the surrogate distribution used in our regularizer. The final training  objective of our model consists of the standard ELBO (Eq. \ref{elbo}) and an additional regularizer, 
$\mathcal L(\theta, \phi) = \mathcal{L}_{ELBO}(\theta, \phi) + \lambda \mathcal{R}(\theta)$.
where $\mathcal{R}(\theta)$ refers to either the calibration regularizer or the reward-based regularizer. Depending on which regularizer is used, it is either minimized together with the negative ELBO or its negative form is used when applying the reward-based regularizer. Note that during training of our model, the regularizer updates only the parameters of the generative model (decoder). The parameters of the surrogate distribution $\hat{p}(\bx|\by)$ remain fixed.
\section{Experiments}
\label{exsperiemnts}
\vspace{-0.4em}

\paragraph{Dataset} We conducted a preliminary investigation of the regularizers' behaviour on the  QM9 dataset \cite{QM9dataset}. We use the octanol-water partition coefficient (LogP) as the conditioning property. Preprocessing details and data analysis are available in Appendix \ref{appndx:dataset}. Additionally, we tested our approach on the ZINC250k dataset, results are presented in Appendix \ref{appndx:zinc}.

\vspace{-1em}
\paragraph{Models} We implemented a baseline conditional VAE along with three versions of the regularizer: the Calibration regularizer, which uses the KL divergence and is referred to as cVAE-KLD, and the two implementations of the reward-based regularizer referred to as cVAE-Pol1 and cVAE-Pol2. 
For each of these, we trained three versions with different decoding strategies (details provided in Appendix \ref{appndx:decoder_structure}): a one-shot decoder, where the sampling of the current token does not depend on the previous token (denoted as cVAE); an auto-regressive decoder, which uses an auto-regressive structure and explicitly samples next tokens at each generation step trained without teacher forcing (EXP-cVAE); and an auto-regressive decoder that employs teacher forcing during training to provide input tokens for the sequence (EXP-cVAE-TF). We use a pre-trained LSTM model as the surrogate. Model architecture and training parameters are given in Appendix \ref{appndx:parameters}.



\vspace{-0.8em}
\paragraph{Evaluation}
Generated sample quality was evaluated using validity, novelty and uniqueness, while conditional performance was assessed with the mean error (MAE) between target and generated properties. Style transfer performance was measured by the proportion of valid style transferred molecules (style transfer validity), the proportion of transfers where the source molecule was reproduced (PCT Fail), Tanimoto similarity of Morgan fingerprints of source and target molecule (Tan) compared with two baselines as lower bounds (Tan-B1 and Tan-B2), and the MAE between target and generated properties. Examples of successful style transfer are provided in Appendix \ref{appndx:style_transfer}. Details on metric computation and sampling are provided in Appendix \ref{appndx:metric_computation}. The code is available on \href{https://github.com/HeathArhturLouis/Rethinking-Molecular-Design-Integrating-Latent-Variable-and-Autoregressive-Models-for-Enhanced-Goal}{GitHub}.

\subsection{Experimental Results}
\label{sec:experimental-results}
\vspace{-0.2em}
\paragraph{One-Shot Decoder Models}
The experimental results for each regularizer type (KLD, Pol1, Pol2) are presented in Tables \ref{one-shot_gen}-\ref{exp-tf-style}. As shown in Tables \ref{one-shot_gen} and \ref{one-shot_style}, models trained with the one-shot version of the decoder perform poorly in generation and conditional generation compared to the baseline cVAE. Both the calibration regularizer (cVAE-KLD) and the reward-based regularizer (cVAE-Pol2) result in reduced validity. Additionally, cVAE-Pol1 suffers from mode collapse. We attribute the failure of the regularizers to the misalignment between the one-shot decoder's molecule generation and the auto-regressive surrogate distribution.

Despite this poor generation performance, the one-shot versions of the models performed better compared to models trained with auto-regressive decoders at the style transfer task (see Table \ref{one-shot_style}), implying latent points and their neighbourhoods belonging to the posterior distribution learned by $q_\phi(\bz|\bx)$ are meaningful under the decoder.

\begin{table}[ht]
\begin{small}
\caption{Conditional generation performance of models trained with one-shot decoders.}
\label{one-shot_gen}
\vspace{0.1in}
\centering
\resizebox{\linewidth}{!}{%
\begin{tabular}{lccccc}
\hline
\textbf{Model} & \textbf{Recon} \boldmath$\uparrow$ & \textbf{Valid} \boldmath$\uparrow$ & \textbf{Uniq} \boldmath$\uparrow$ & \textbf{Novelty} \boldmath$\uparrow$ & \textbf{Prop MAE} \boldmath$\downarrow$ \\ \hline
 LSTM  &-& 0.9700 & 1.0000 & 0.3024 & 0.3211 \\
         \hline
cVAE & 0.8872 & 0.2210 & 1.0000 & 0.8959 & 0.7976 \\
\textbf{cVAE-KLD} & 0.8796 & 0.0110 & 1.0000 & 0.9090 & 0.7919 \\
\textbf{cVAE-Pol1} & 0.9436 & 1.0000 & 0.0010 & 0.0000 & 1.0121 \\
\textbf{cVAE-Pol2} & 0.8769 & 0.0050 & 1.0000 & 1.0000 & 1.4675 \\
\hline 
\cite{DBLP:journals/corr/Gomez-Bombarelli16}& 0.0361 &0.1030 &- &0.9000&-\\
\hline
\end{tabular}
}
\vspace{-1em}
\caption{Style transfer performance of cVAE with one-shot decoders.}
\label{one-shot_style}
\vspace{0.1in}
\resizebox{\linewidth}{!}{%
\begin{tabular}{lllllll}
\hline
\textbf{Model} & \textbf{PCT Valid \boldmath$\uparrow$} & \textbf{PCT Fail \boldmath$\downarrow$} & \textbf{MAE} \boldmath$\downarrow$ & \textbf{Tan} \boldmath$\uparrow$ & \textbf{Tan-B1} & \textbf{Tan-B2} \\ \hline

cVAE &  0.9260 & 0.7482 & 0.8636 & 0.4194 & 0.0622 & 0.0655 \\
\textbf{cVAE-KLD} & 0.9072 & 0.7579 & 0.8395 & 0.3894 & 0.0490 & 0.0677 \\
\textbf{cVAE-Pol1} & 0.9468 & 0.8906 & 0.9909 & 0.4451 & 0.0296 & 0.0652 \\
\textbf{cVAE-Pol2} & 0.9148 & 0.7420 & 0.8497 & 0.4111 & 0.0536 & 0.0659 \\
\hline
\end{tabular}
}
\vspace{-1em}
\end{small}
\end{table}
\vspace{-0.3em}
\paragraph{Auto-Regressive Decoder Models} To ensure alignment with the surrogate distribution used in the regularizer, which is trained auto-regressively, we 
also investigated the auto-regressive decoder models. However, to prevent latent representation degradation due to the strong decoder, we abstained from using teacher forcing. Instead, at each step, we sampled from the model's own prediction, forcing the decoder to rely on the latent representation during generation. The results are presented in Table \ref{exp-gen} and \ref{exp-style}. Tables demonstrates promising results for the calibration regularizer (EXP-cVAE-KLD) and reward-based regularzier (EXP-cVAE-Pol2). Compared to the baseline, both improve the validity of the generated molecules and have significant improvement in conditonal generation as well as the style transfer task. However, the  EXP-cVAE-Pol1 model continues to fail. 

\begin{table}[ht]
\begin{small}
\caption{Conditional generation performance of models trained with auto-regressive decoders without teacher forcing.}
\label{exp-gen}
\vspace{0.1in}
\centering
\resizebox{\linewidth}{!}{%
\begin{tabular}{lccccc}
\hline
\textbf{Model} & \textbf{Recon} \boldmath$\uparrow$ & \textbf{Valid} \boldmath$\uparrow$ & \textbf{Uniq} \boldmath$\uparrow$ & \textbf{Novelty} \boldmath$\uparrow$ & \textbf{Prop MAE} \boldmath$\downarrow$ \\ \hline
EXP-cVAE  & 0.9509 & 0.2860 & 1.0000 & 0.8287 & 0.7058 \\
\textbf{EXP-cVAE-KLD} & 0.6783 & 0.7460 & 1.0000 & 0.5161 & 0.2831 \\
\textbf{EXP-cVAE-Pol1} & 0.9363 & 0.0010 & 1.0000 & 1.0000 & 0.1987 \\
\textbf{EXP-cVAE-Pol2} & 0.6711 & 0.8250 & 0.9976 & 0.4642 & 0.2300 \\
\hline
\end{tabular}
}
\vspace{-0.5em}
\caption{Style transfer performance of cVAE with auto-regressive decoder trained without teacher forcing.}
\label{exp-style}
\vspace{0.1in}
\resizebox{\linewidth}{!}{%
\begin{tabular}{lllllll}
\hline
\textbf{Model} & \textbf{PCT Valid.} \boldmath$\uparrow$ & \textbf{PCT Fail} \boldmath$\downarrow$ & \textbf{MAE} \boldmath$\downarrow$ & \textbf{Tan} \boldmath$\uparrow$ & \textbf{Tan-B1} & \textbf{Tan-B2} \\ \hline
EXP-cVAE & 0.8968 & 0.7243 & 0.7514 & 0.3711 & 0.0653 & 0.0646 \\
\textbf{EXP-cVAE-KLD}  & 0.8260 & 0.4459 & 0.5377 & 0.3507 & 0.0679 & 0.0667 \\
\textbf{EXP-cVAE-Pol1} & 0.8888 & 0.7115 & 1.5314 & 0.3206 & 0.0209 & 0.0666 \\
\textbf{EXP-cVAE-Pol2} & 0.8404 & 0.4131 & 0.5493 & 0.3481 & 0.0653 & 0.0662 \\
\hline
\end{tabular}
}
\vspace{-1em}
\end{small}
\end{table}
\vspace{-0.5em}

In our final exploration, we extended the decoder to full auto-regressive setting trained with teacher forcing (Tables \ref{exp-tf-gen}, \ref{exp-tf-style}). As expected, the introduction of teacher forcing significantly improves the decoder's ability to generate valid molecules. Consistent as previous results, the two regularizer (EXP-cVAE-TF-KLD and EXP-cVAE-TF-Pol2) greatly improves both the proportion of valid molecules and the conditional performance of the model. In terms of style transfer performance we see a similar trend to the non teacher forcing case, with the regularized version of the model having a much lower proportion of failed style transfer attempts. To illustrate this we have included a plot of the style transfer performance of both models (EXP-cVAE-TF-KLD and EXP-cVAE-TF) in Appendix \ref{appndx:style_transfer}. 

Note that the results of the auto-regressive decoder (trained with teacher forcing) contradict the common assumption that a ``strong decoder'' could degrade the latent space. Therefore, we conducted a more in-depth analysis, which we provide in Appendix Section \ref{strong_decoder}. We conclude when trained in an auto-regressive fashion with teacher forcing, the latent representation is utilized until a certain length of the sequence, beyond which the decoder does not rely on the latent representation to generate the rest of the sequence.





\begin{table}[ht]
\begin{small}
\caption{Conditional generation performance of models trained with auto-regressive decoders with teacher forcing.}
\label{exp-tf-gen}
\vspace{0.1in}
\centering
\resizebox{\linewidth}{!}{%
\begin{tabular}{lccccc}
\hline
\textbf{Model} & \textbf{Recon} \boldmath$\uparrow$ & \textbf{Valid} \boldmath$\uparrow$ & \textbf{Uniq} \boldmath$\uparrow$ & \textbf{Novelty} \boldmath$\uparrow$ & \textbf{Prop MAE} \boldmath$\downarrow$ \\ \hline
EXP-cVAE-TF & 0.9481 & 0.8340 & 1.0000 & 0.7014 & 0.3636 \\
\textbf{EXP-cVAE-TF-KLD } & 0.6857 & 0.9580 & 0.9958 & 0.3862 & 0.1292 \\
\textbf{EXP-cVAE-TF-Pol1} & 0.9575 & 0.0010 & 1.0000 & 0.0000 & 0.0816 \\
\textbf{EXP-cVAE-TF-Pol2} & 0.6799 & 0.9710 & 0.9938 & 0.3326 & 0.1259 \\
\hline
Lim et. al.-TF & 0.4381 & 0.5700 & 0.9863 & 0.9520 & 0.4510 \\
\hline
\end{tabular}
}
\vspace{-1em}

\caption{Style transfer performance of cVAE with auto-regressive decoder trained with teacher forcing.}
\label{exp-tf-style}
\vspace{0.1in}
\resizebox{\linewidth}{!}{%
\begin{tabular}{lllllll}
\hline
\textbf{Model} & \textbf{PCT Valid} \boldmath$\uparrow$ & \textbf{PCT Fail} \boldmath$\downarrow$ & \textbf{MAE} \boldmath$\downarrow$ & \textbf{Tan} \boldmath$\uparrow$ & \textbf{Tan B1} & \textbf{Tan B2} \\ \hline
EXP-cVAE-TF & 0.9176 & 0.3391 & 0.3955 & 0.3071 & 0.0667 &  0.0668 \\
\textbf{EXP-cVAE-TF-KLD } & 0.9668 & 0.0852 & 0.1493 & 0.2328 & 0.0651 & 0.0658 \\
\textbf{EXP-cVAE-TF-Pol1} & 0.4312 & 0.6790 & 0.2935 & 0.3614 & 0.0532 & 0.0652 \\
\textbf{EXP-cVAE-TF-Pol2 } & 0.9656 & 0.0849 & 0.1486 & 0.2280 & 0.0652 & 0.0651 \\
\hline
Lim et al.-TF & - & 0.0210 & 0.3439 & 0.1424 & 0.08160 & 0.11450 \\
\hline
\end{tabular}
}
\vspace{-1em}
\end{small}
\end{table}

\vspace{-0.5em}
\section{Conclusions}
Our framework addresses key limitations in classic generative methods, particularly enhancing conditional generation and style transfer. Novel regularizers, such as the calibration and reward-based regularizers, guide the generative process, improving the fidelity and validity of generated molecules. This work is intended to return to fundamental models based on simple SMILES representations to investigate commonly accepted hypotheses and provide in-depth analysis. 
Experimental results demonstrated the potential of the proposed regularizers, however, further investigation and testing are still needed. Our work also provided insights about the models and assumption used in molecular design, paving the way for future research in more complex settings.

\newpage
\nocite{langley00}
\bibliography{main}
\bibliographystyle{icml2024}

\newpage
\appendix
\onecolumn

\section{Literature Review}
\label{appendx:lit_review}
\subsection{Auto Regressive Models Trained on SMILES}

Early auto-regressive approaches to molecule generation \cite{segler2017generating} \cite{Bjerrum2017MolecularGW} showed that auto-regressive model can successfully learn a distribution of SMILES strings and generate realistic candidates with high accuracy. Furthermore, \citet{segler2017generating} proposed conditional generation be achieved by filtering the generated molecules for candidates with desirable properties or by fine tuning the model on a curated dataset of desirable molecules. Methods have also been developed to directly control the conditioning and avoid the need for further fine-tuning. For example, concatenating conditional information to inputs at the first or every generation step and encoding conditional information into the model's initial hidden state \cite{Kotsias2020DirectSO}. 

A parallel approach to guiding auto-regressive models for conditional generation uses policy optimization and a set of objective functions to fine-tune auto retrogressive models. The seminal examples in this direction are REINVENT \cite{Olivecrona2017MolecularDD} and ReLeaSe \cite{2017DeepRL}. More recent work has incorporated various other innovations from the field of reinforcement learning such as \cite{Thiede2020CuriosityIE} and curriculum learning \cite{Mokaya2022TestingTL}.

Following their success in the field of natural language processing, the majority of recent publications have adopted various transformer based architectures. For example, \citet{Wang2019LearningDT} and \citet{Grechishnikova2019TransformerNN} propose transformer based approaches to structure based drug design. Another notable recent example, \citet{Wang2021MulticonstraintMG} propose a multi-step pipeline where knowledge is distilled from a transformer model to an RNN-based model which is then further fine-tuned using reinforcement learning. 

Although these have been found to outperform older architectures such as RNN and LSTM, the difference is minimal in simpler settings \cite{Bagal2021MolGPTMG}. For this reason and to maintain a similar architecture between the VAE decoder and auto-regressive model and prior VAE based models, we have opted to employ a conditional LSTM model in this study.

\subsection{Latent Variable Models Trained on SMILES}
The seminal implementation of a latent variable model for SMILES \cite{DBLP:journals/corr/Gomez-Bombarelli16} is based on a conditional character VAE that encodes SMILES as continuous latent vectors. It also uses a network to predict molecular properties from latent vectors, allowing for efficient property optimization. \citet{Lim2018MolecularGM} propose an approach based on a conditional VAE for simultaneous control of multiple properties. 
Although successful, the SMILES generated by the model are prone to syntactic and semantic errors. Following works aimed to address this by operating on grammar rules for generating SMILES \cite{kusner2017grammar}, and later on attribute grammars \cite{dai2018syntax, mollaysa2019disentanagled} (to address errors that are still possible when using a regular CFG). These models were able to generate a much higher proportion of valid SMILES at the cost of computational and implementation complexity. 

Others works such as \citet{simonovsky2018graphvae}, \citet{jin2018junction}, \citet{you2018graph} and \citet{liu2018constrained} explored models that operate on graph representations, largely improving the ability to generate valid molecules. A notable example is JT-VAE \cite{jin2018junction} witch is able to guarantee validity of generated molecules by constructing molecules from a library of fragments.


\subsection{Auto-Regressive Models with Latent Variables}

There have been several attempts to introduce latent variables into RNNs such VRNN \cite{chung2015recurrent}, and others \cite{bayer2014learning, boulanger2012modeling, fabius2014variational}. However, the goal of such latent variables is to
introduce more stochasticity into the model as a standard RNN has limited power to model
the variability observed in highly-structured data. Integrating latent variables into the hidden state of the RNN aims to be able to
model more variability that is observed in the structured data. When used to model the
conditional distribution $p(\bx|\by)$, while such models help to learn a distribution with more diverse samples compared to standard RNN models, they do not provide sufficient control over the style of the generated samples.
Recently, an approach similar to VRNN, \citet{bird2019customizing} also introduces a latent variable $\bz$ into
RNNs, in addition to modelling more variability of sequences, this paper aims to capture sequence-specific features through latent the
variable $\bz$ and use it to perform style transfer. The latent variable $\bz$ is inferred from input output sequence pairs and the parameters of the RNN are made dependent on $\bz$. This
enables each sequence to be modelled as a single dynamical system and capture instance-specific
variation through $\bz$ which later helps us to perform style transfer over sequences.
Due to auto-regressive models' strong conditional generation performance, it would
be interesting to perform style transfer with auto-regressive models in an unsupervised
fashion without the need of constructing a paired dataset.


\subsection{Strong Decoder}
\label{strong_decoder}

It has been assumed or stated by \cite{DBLP:journals/corr/Gomez-Bombarelli16} that when trained with teacher forcing, the auto-regressive decoder tends to ignore the encoded $\bz$ and recover the next token purely from the current token. However, this assumption was overlooked by \cite{lim2018molecular}, who proceeded with a similar type of model featuring a strong decoder (an RNN trained with teacher forcing). Mathematically, it is reasonable to assume that if one can learn $p(\bx|\by)$ using an auto-regressive model trained to predict the next token \cite{segler2017generating}, then when training a generative model $p(\bx|\bz, \by)$ through a VAE using the same model as a decoder, the generative model could easily converge to the auto-regressive model $p(\bx|\by)$ by ignoring $\bz$. This is because it is easy to recover the next token $p(x_t|\by, x_1, \dots, x_{t-1}, \bz)$ without relying on $\bz$ at all.
This assumption has been generally accepted in subsequent research papers, which have trained $p(\bx|\bz, \by)$ without the auto-regressive component when using latent variable models to avoid degenerated latent representations.

To investigate the impact of the latent variable $\bz$, we took the model from \cite{lim2018molecular} and tested if their model leads to degraded latent representation. The first reliable measure would be reconstruction, but they did not report the reconstruction performance of their model. Since the model training is very slow, we retrained their model on the QM9 dataset as a proof of concept.

The model performance is reported in Tables \ref{exp-tf-gen} and \ref{exp-tf-style}. It shows that we have a reconstruction rate of 0.43. Note that $\by$ also carries some information about $\bx$. Therefore, to test further, we tried the reconstruction with latent $\bz$ that is sampled from the prior instead of sampled from the posterior. This resulted in a 0\% reconstruction rate, which further validates that the latent variable $\bz$ does carry useful information about the encoded $\bx$.

After investigating whether the learned latent variable $\bz$ carries meaningful information, we discovered that it does, but only up to a certain step in the auto-regressive generation process. For instance, with the QM9 dataset, where the maximum sequence length T is under 40, the decoder operates in an auto-regressive setting. During reconstruction, while decoding progressively, we observed the following results presented in Table \ref{lim_et_all_z}:

\begin{table}
\caption{Reconstruction rates with different conditions.}
\label{lim_et_all_z}
\vskip 0.1in
\centering
\begin{tabular}{ll}
\hline
\textbf{Condition} & \textbf{Reconstruction} \\
\hline
Standard (using $\bz$ throughout) & 0.4351 \\
Using $\bz$ up to 15th step & 0.4312 \\
Using $\bz$ up to 10th step & 0.3922 \\
Using $\bz$ up to 1st step & 0.0025 \\
Not using $\bz$ (prior from step zero) & 0.0000 \\
\hline
\end{tabular}

\end{table}
This indicates that $\bz$ carries information that is utilized only until the 16th step, after which it becomes ineffective. This behavior arises from the auto-regressive training with teacher forcing. Initially, when generating the next token conditioned on $\by$ and the current token, there are many possible characters that could be the next token, each leading to a SMILES string with the property $\by$. Therefore, it is crucial for $\bz$ to guide the decoding process to ensure the generation of the specific $\bx$ that was used to encode $\bz$. This makes $\bz$ essential for generating the next token at the beginning of the sequence. However, as the sequence progresses, the number of possible next tokens significantly reduces, making it clear what the next token should be based on the current token. Consequently, $\bz$ becomes less useful, leading the model to utilize $\bz$ primarily for encoding the initial structure of the original SMILES string, but not for the entire sequence.


\subsection{Reward-Based Regularizer Behaviour}

The initial version of the reward-based regularizer (Pol1) failed to produce a high proportion of valid and unique molecules, regardless of decoder stricture. This may be due to high variance during training as a result of relying on the score function gradient estimate.  We attempted to fix this by introducing the regularizer at later points in the training process (where the decoder has learned to generate better molecules) and by testing various decoder scheduling strategies with little success.

\subsection{Analysis}

\paragraph{Can we actually disentangle the latent variable $\bz$ and property $\by$? }
First of all, unlike in the space of image where we can actually annotate the variations in the
data, molecular space is much more complicated. When we apply vanilla VAE to encode
molecules to a latent space, we can assume that $\bz$ encodes the molecular structure, and previous works have shown that points that are close in the latent space result in molecules
that are structurally similar. However, once we assume a molecule is generated from a property variable y and some other latent factor $\bz$ where $\by$ and $\bz$ are independent under the
prior, $p(\bx, \by, \bz) = p(\bx|\by, \bz)p(\bz)p(\by)$, then we can not really say what $\bz$ encodes. Therefore,
it is not really clear what $\bz$ encodes in this setting.
 Even though we still assume $\bz$ encodes some structural features of the molecules, for instance, some functional groups that are not dependent on the property $\by$, it is hard to
clarify what aspect of the molecule that $\bz$ learns.
However, it is clear is that $\bz$ does not fully specify the molecule’s structure since the property
of the molecule is fully determined by the structure of the molecule, and we assume the
learned $\bz$ is disentangled from $\by$. We can not modify the property of a molecule without
modifying its structure.
 However, note that $\bz$ is a learned
factor, assuming that it is independent of the $\by$. Therefore, $\bz$ should not learn some variations in the molecule data that is independent of the conditioning properties.  However, in reality, for a given molecule, it is not clear that if 
we can always modify its property to the value we want and stay close to the original molecules
in terms of structure \cite{aminanmu2021structural}.








\newpage
\section{Additional Experiments With the ZINC250k Dataset}
\label{appndx:zinc}

Since submitting this work for review, we have run additional experiments on the ZINC250k subset of the ``ZINC is not commercial'' dataset \cite{irwin2005zinc}. The ZINC250k dataset contains more SMILES that are individually longer and more complex. Additionally all compounds are commercially available and highly drug-like. This makes ZINC a more realistic setting than the enumerated molecular structures in QM9. 

\begin{figure}[h]
    \centering
    \includegraphics[width=0.8\linewidth]{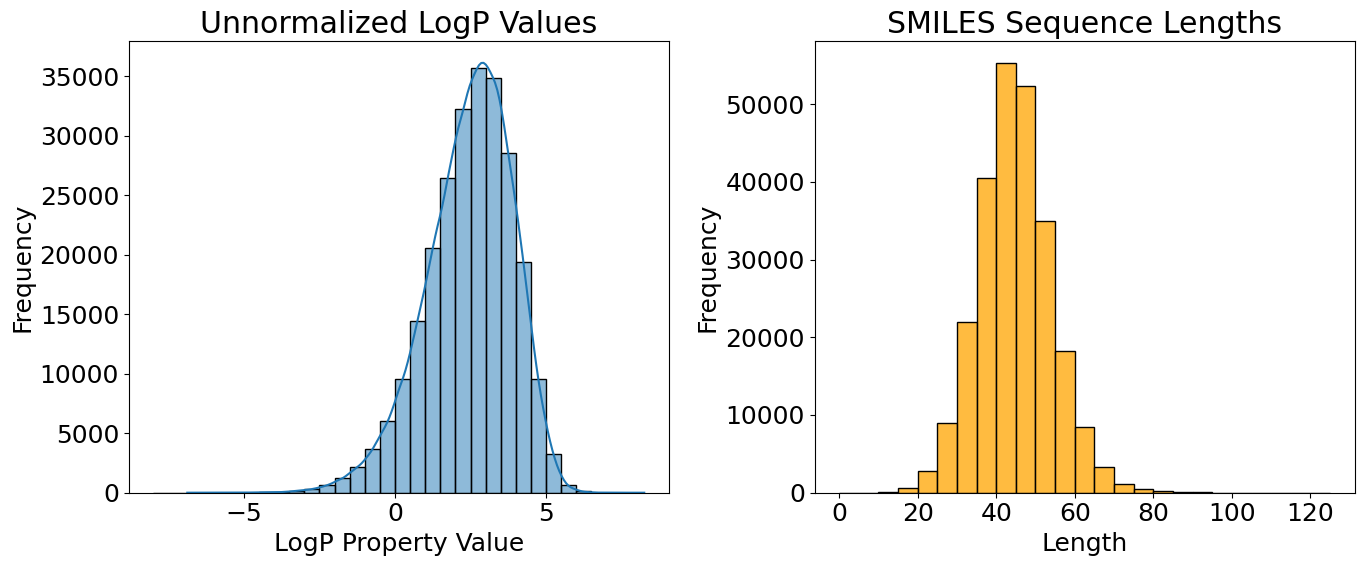}
    \caption{Summary statistics for ZINC250k dataset.}
    \label{fig:ZINC-sum-stats}
\end{figure}

ZINC SMILES were pre-processed with the same numerical encoding scheme used for QM9, with sequences padded to a final length of 120. The baseline upper-bound MAE between random pairs of test properties is $1.6067$. Summary statistics of the sequences and properties are presented in Figure \ref{fig:ZINC-sum-stats} and the distribution of SMILES characters in Table \ref{tab:ZINC250k_char_freqs}.

\begin{table}[ht]
\caption{SMILES symbol frequencies in the ZINC250k dataset.}
\label{tab:ZINC250k_char_freqs}
\vskip 0.15in
\centering
\resizebox{\linewidth}{!}{%
\begin{minipage}[t]{0.32\linewidth}
\centering
\begin{tabular}{ccc}
\hline
\textbf{Char.} & \textbf{Freq.} & \textbf{Norm. Freq.} \\
\hline
c & 2208591 & 0.20 \\
C & 2089152 & 0.19 \\
( & 963303 & 0.09 \\
) & 963303 & 0.09 \\
1 & 724766 & 0.07 \\
O & 545244 & 0.05 \\
2 & 471857 & 0.04 \\
N & 437084 & 0.04 \\
= & 395693 & 0.04 \\
n & 268119 & 0.02 \\
3 & 166892 & 0.02 \\
{[} & 344325 & 0.03 \\
\hline
\end{tabular}
\end{minipage}%
\hfill%
\begin{minipage}[t]{0.32\linewidth}
\centering
\begin{tabular}{ccc}
\hline
\textbf{Char.} & \textbf{Freq.} & \textbf{Norm. Freq.} \\
\hline
{]} & 344325 & 0.03 \\
@ & 335552 & 0.03 \\
H & 290126 & 0.03 \\
F & 79430 & 0.01 \\
+ & 76813 & 0.01 \\
S & 63791 & 0.01 \\
- & 63474 & 0.01 \\
l & 42872 & 0.00 \\
s & 39088 & 0.00 \\
o & 30987 & 0.00 \\
4 & 30588 & 0.00 \\
/ & 28557 & 0.00 \\
\hline
\end{tabular}
\end{minipage}%
\hfill%
\begin{minipage}[t]{0.32\linewidth}
\centering
\begin{tabular}{ccc}
\hline
\textbf{Char.} & \textbf{Freq.} & \textbf{Norm. Freq.} \\
\hline
\# & 15587 & 0.00 \\
r & 12722 & 0.00 \\
B & 12722 & 0.00 \\
\textbackslash & 5839 & 0.00 \\
5 & 2592 & 0.00 \\
I & 888 & 0.00 \\
P & 127 & 0.00 \\
6 & 88 & 0.00 \\
7 & 8 & 0.00 \\
8 & 2 & 0.00 \\
& & \\
& & \\
\hline
\end{tabular}
\end{minipage}
}
\end{table}

\subsection{ZINC250k Results and Discussion}

We retrained base models with each decoder type and all working configurations of the regularizers. The results for these are presented in Tables \ref{tab:zinc-gen}, \ref{tab:zinc-st-transfer}. During training we used a surrogate model with the architecture described in Appendix Section \ref{appndx:parameters} whose performance is presented in Table \ref{tab:zinc-surrogate}. We note that the novelty issue observed with the surrogate trained on QM9 is not present in ZINC250k. As predicted, the introduction of either regularizer (Table \ref{tab:zinc-gen}) does not decrease novelty. Additionally, the one-shot decoder model struggled to generate a non-trivial proportion of valid SMILES, in line with the experience of \cite{DBLP:journals/corr/Gomez-Bombarelli16}.

\begin{table}[ht] 
\caption{Performance of surrogate LSTM model for ZINC250k.}
\label{tab:zinc-surrogate}
\vskip 0.1in
\centering
\begin{small}
\resizebox{0.5\linewidth}{!}{
\begin{tabular}{lcccc}
\hline
\textbf{Model} & \textbf{Valid \boldmath$\uparrow$} & \textbf{Uniq \boldmath$\uparrow$} & \textbf{Novelty \boldmath$\uparrow$} & \textbf{Prop MAE \boldmath$\downarrow$} \\
\hline
\textbf{LSTM} & 0.9333 & 1.0000 & 0.9964 & 0.1714 \\
\hline
\end{tabular}
}
\end{small}
\end{table} 

We see that our observations from the QM9 dataset carry over to the ZINC scenario. Conditional generation performance (Table \ref{tab:zinc-gen}) is improved both in terms of validity and property error. These changes are particularly dramatic in the case of the explicit decoder, where the model performs extremely poorly in both cases. The effect on style transfer performance (Table \ref{tab:zinc-st-transfer}) of models' trained with the regularizer is also similar. Introduction of either form of the regularizer improves the target-generation MAE and decreases the proportion of style transfer failures. Additionally, it slightly decreases the source-generated structural similarity in the explicit case. As with QM9, the Pol2 regularizer appears to have a stronger effect on both tasks, with the exception of the proportion of style transfer failures which is surprisingly higher. 

Although the direction of the changes introduced by the regularizers is the same, the absolute performance levels of the models lead to a different trade-off. One difference from the QM9 case, is that the teacher forcing decoder is less effective by itself, only achieving a validity of $0.568$ compared to $0.938$ when trained with the Pol2 regularizer. Additionally, in the QM9 case, we see that without teacher forcing, the decoder puts significant emphasis on $\bz$, and therefore achieves a poor MAE even when the regularizer is used. Overall, these differences make the regularized teacher forcing model appear attractive relative to the explicit one.

\begin{table}[ht] 
\caption{Conditional generation performance of ZINC250k models.}
\label{tab:zinc-gen}
\vskip 0.1in
\centering
\begin{small}
\resizebox{0.7\linewidth}{!}{
\begin{tabular}{lccccc}
\hline
\textbf{Model} & \textbf{Recon \boldmath$\uparrow$} & \textbf{Valid \boldmath$\uparrow$} & \textbf{Uniq \boldmath$\uparrow$} & \textbf{Novelty \boldmath$\uparrow$} & \textbf{Prop MAE \boldmath$\downarrow$} \\
\hline
\textbf{cVAE} & 0.1852 & 0.0010 & 1.0000 & 1.0000 & 1.0228 \\
\hline
\textbf{EXP-cVAE} & 0.4301 & 0.0020 & 1.0000 & 1.0000 & 1.5033 \\
\textbf{EXP-cVAE-KLD} & 0.1716 & 0.8550 & 1.0000 & 1.0000 & 0.3690 \\
\textbf{EXP-cVAE-Pol2} & 0.4321 & 0.9360 & 1.0000 & 1.0000 & 0.2065 \\
\hline
\textbf{EXP-cVAE-TF} & 0.5832 & 0.5680 & 1.0000 & 1.0000 & 0.6600 \\
\textbf{EXP-cVAE-TF-KLD} & 0.3439 & 0.9110 & 1.0000 & 0.9989 & 0.3283 \\
\textbf{EXP-cVAE-TF-Pol2} & 0.5369 & 0.9380 & 1.0000 & 0.9989 & 0.1998 \\
\hline
\end{tabular}
}
\vspace{-0.5em}
\end{small}
\end{table} 

\begin{table}[ht] 
\caption{Style transfer performance of ZINC250k models.}
\label{tab:zinc-st-transfer}
\vskip 0.1in
\centering
\begin{small}
\resizebox{0.7\linewidth}{!}{
\begin{tabular}{lcccccc}
\hline
\textbf{Model} & \textbf{PCT Valid \boldmath$\uparrow$} & \textbf{PCT Fail \boldmath$\downarrow$} & \textbf{MAE \boldmath$\downarrow$} & \textbf{Tan \boldmath$\uparrow$} & \textbf{Tan-B1} & \textbf{Tan-B2} \\
\hline
\textbf{cVAE} & 0.3944 & 0.3765 & 1.5439 & 0.5755 & 0.0921 & 0.1186 \\
\hline
\textbf{EXP-cVAE} & 0.6492 & 0.5699 & 1.3499 & 0.6227 & 0.0741 & 0.1164 \\
\textbf{EXP-cVAE-KLD} & 0.4308 & 0.2953 & 1.3105 & 0.5700 & 0.1125 & 0.1149 \\
\textbf{EXP-cVAE-Pol2} & 0.6692 & 0.5642 & 1.2572 & 0.6115 & 0.1138 & 0.1164 \\
\hline
\textbf{EXP-cVAE-TF} & 0.8448 & 0.2656 & 0.7114 & 0.4311 & 0.1013 & 0.1136 \\
\textbf{EXP-cVAE-TF-KLD} & 0.8456 & 0.0941 & 0.4119 & 0.4298 & 0.1129 & 0.1115 \\
\textbf{EXP-cVAE-TF-Pol2} & 0.8724 & 0.1444 & 0.3759 & 0.4214 & 0.1145 & 0.1142 \\
\hline
\end{tabular}
}
\end{small}
\end{table} 

\newpage
\section{Examples of Style Transfer Molecules}
\label{appndx:style_transfer}

\begin{figure}[H]
    \centering
    \includegraphics[width=1.0\linewidth]{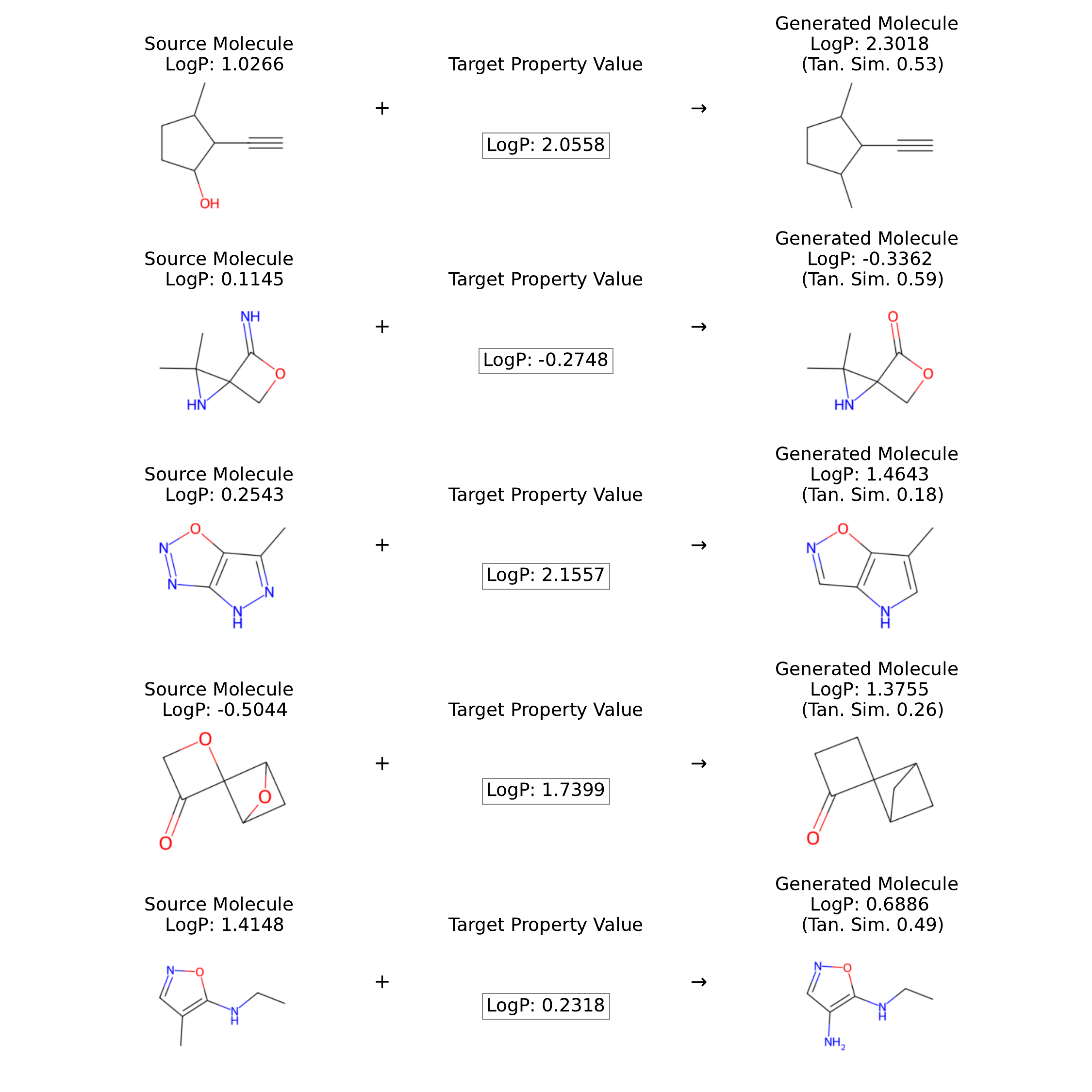}
    \caption{Examples of successful style transfer with EXP-cVAE-TF-Pol2 model.}
    \label{fig:KLD_ST}
\end{figure}

Figure \ref{fig:KLD_ST} shows five examples of successful style transfer for a model trained with the calibration regularizer and explicit decoder (without teacher forcing). We see the model is able to find relatively simple and effective solutions to alter the LogP value.

\begin{figure}[H]
    \centering
    \includegraphics[width=\linewidth]{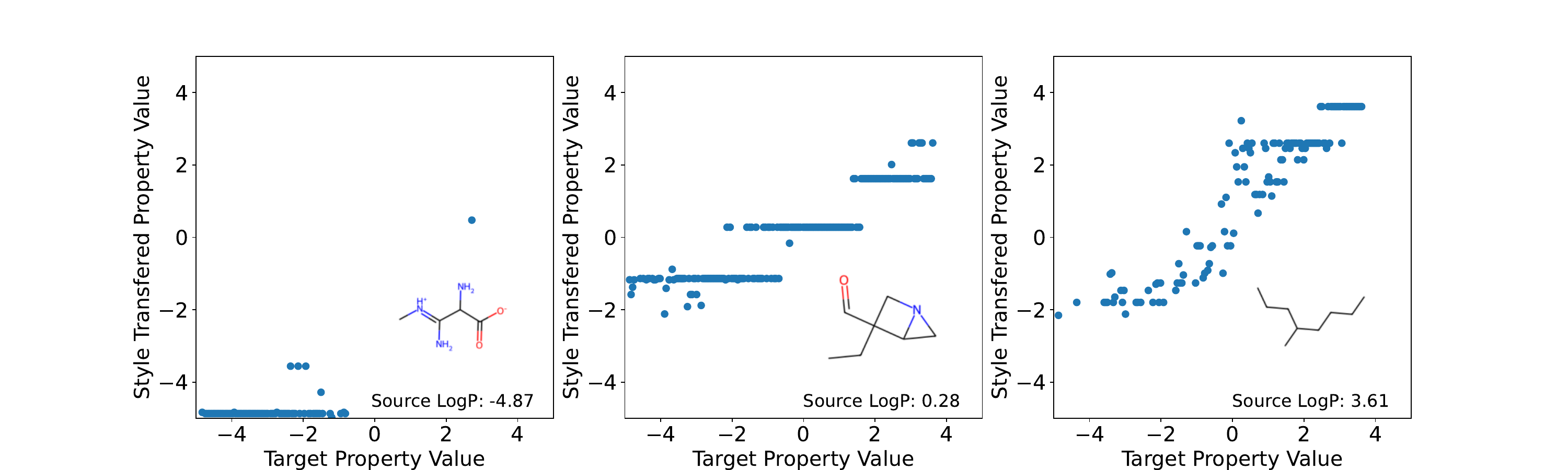}
    \caption{Style transfer generation property scores for a range of target properties sampled from EXP-cVAE model.}
    \label{fig:Range_cVAE}
\end{figure}

\begin{figure}[H]
    \centering
    \includegraphics[width=\linewidth]{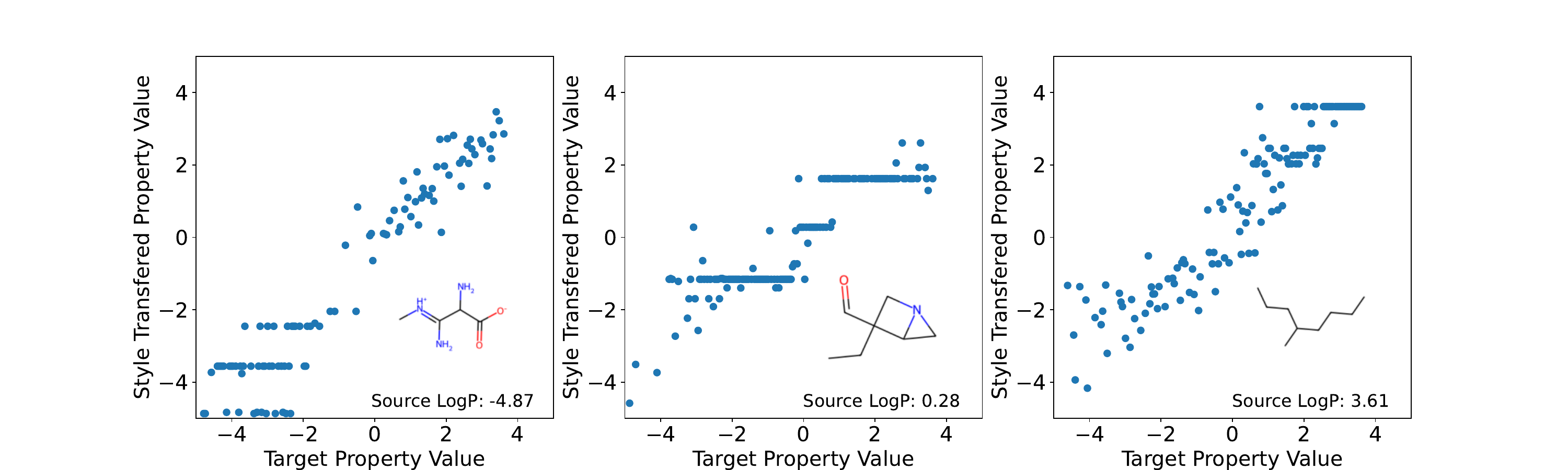}
    \caption{Style transfer generation property scores for a range of target properties sampled from EXP-cVAE-KLD model.}
    \label{fig:Range_KLD}
\end{figure}

\begin{figure}[H]
    \centering
    \includegraphics[width=\linewidth]{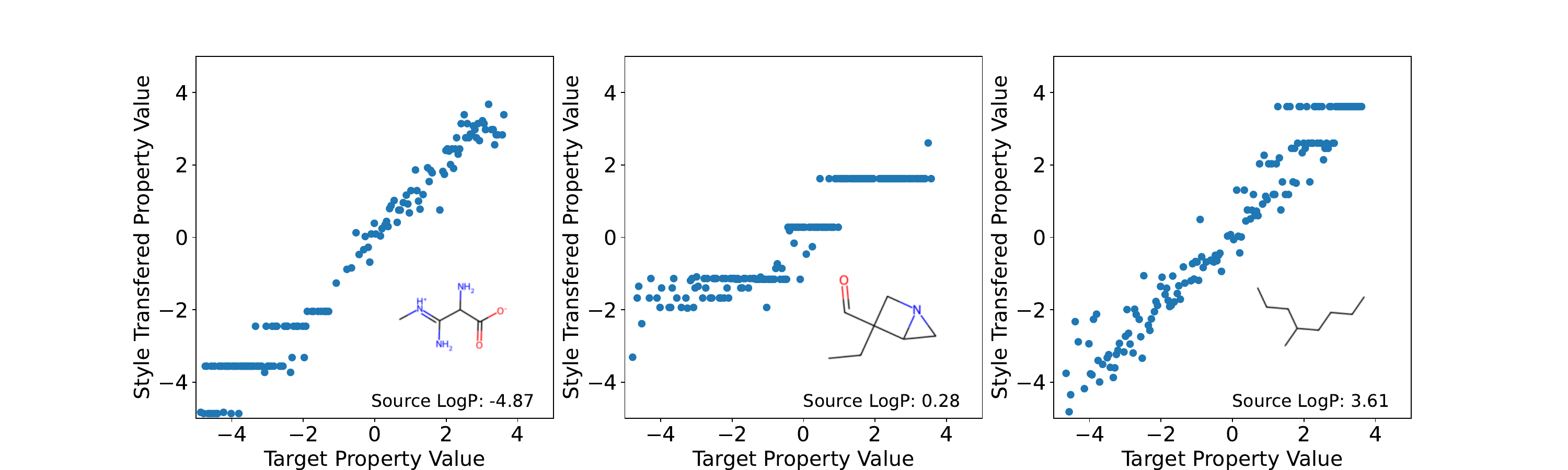}
    \caption{Style transfer generation property scores for a range of target properties sampled from EXP-cVAE-Pol2 model.}
    \label{fig:Range_Pol2}
\end{figure}

To better visualise the positive effect of our regularizers, we can examine their ability to edit a molecular structure for a range of target property scores. To show this, we plotted the achieved property scores over the target property scores for the molecules with the lowest, median and highest LogP scores. These results are presented for the explicit decoder models with no regularizer (Figure \ref{fig:Range_cVAE}), calibration regularizer (Figure \ref{fig:Range_KLD}), and the second reward regularizer (Figure \ref{fig:Range_Pol2}). 

An ideal model would perfectly match the target and contain points along the diagonal. We see that the introduction of the regularizers (bottom two plots) are significantly closer to this. In particular they more flexible, and are able to find more distinct solution modes. These are visible in the horizontal striations. The uneven structure of these plots demonstrates the dependence on the ability of a model to perform style transfer on both the source molecule and target property. For example, the molecule on the right hand side appears to be easier to edit for all models. 

\begin{figure}[H]
    \centering
    \includegraphics[width=\linewidth]{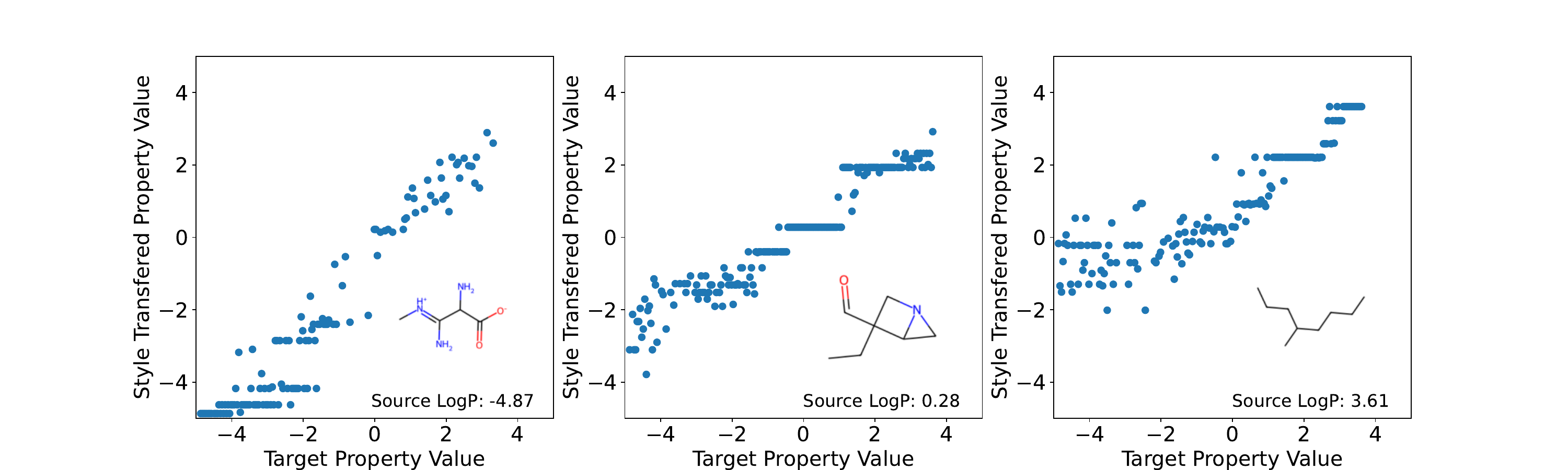}
    \caption{Style transfer generation property scores for a range of target properties sampled from EXP-cVAE-TF model.}
    \label{fig:TF-Range_cVAE}
\end{figure}

\begin{figure}[H]
    \centering
    \includegraphics[width=\linewidth]{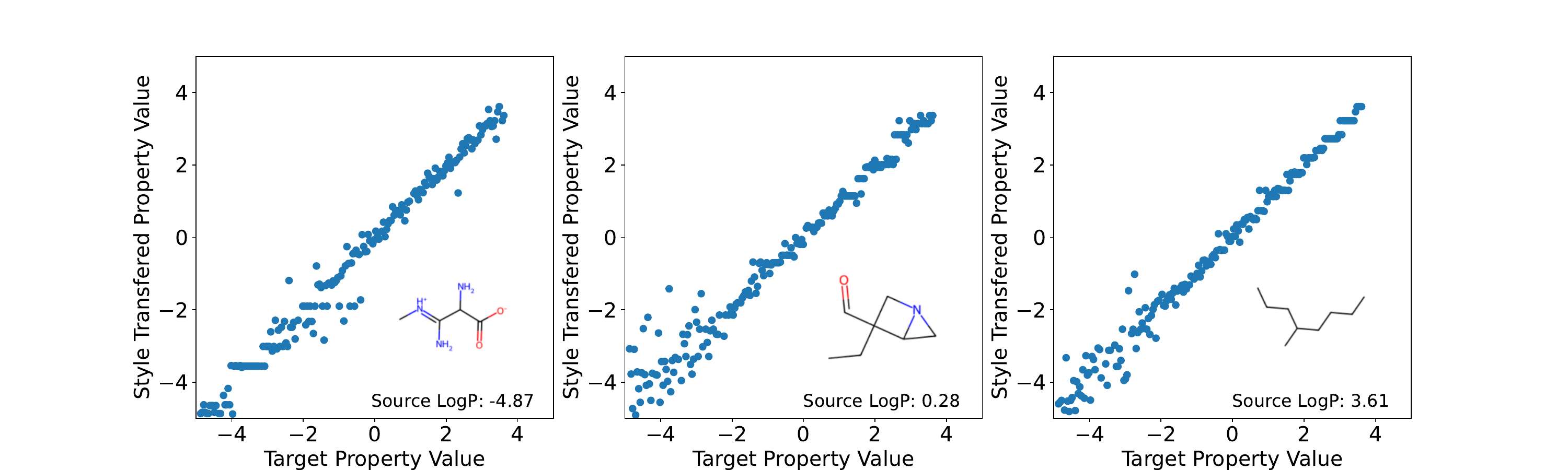}
    \caption{Style transfer generation property scores for a range of target properties sampled from EXP-cVAE-TF-KLD model.}
    \label{fig:TF-Range_KLD}
\end{figure}

\begin{figure}[H]
    \centering
    \includegraphics[width=\linewidth]{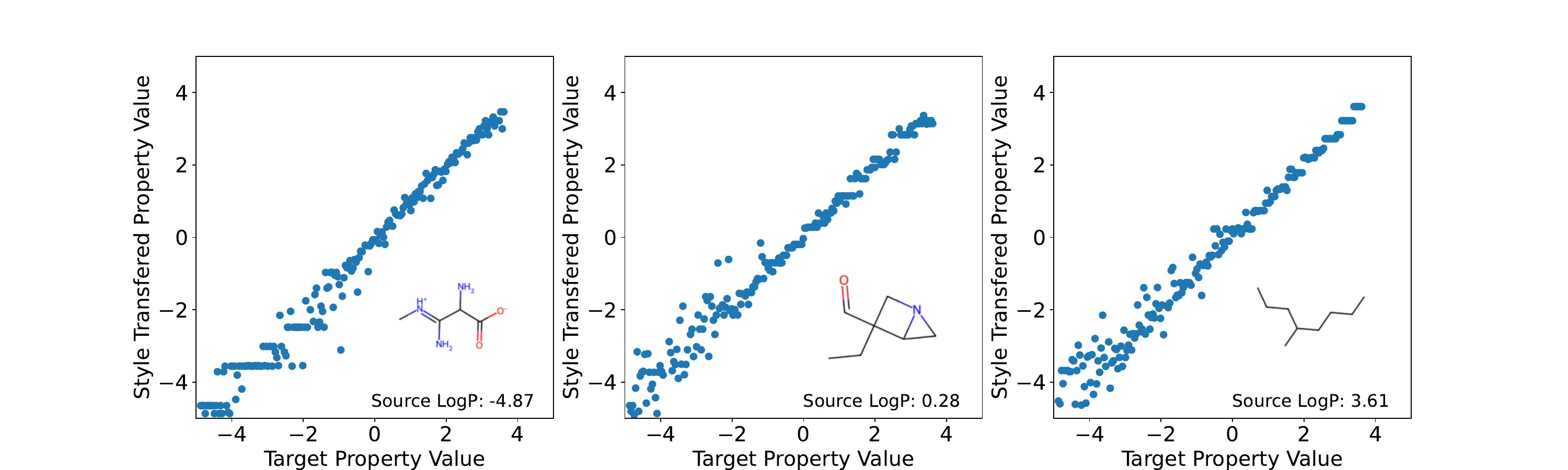}
    \caption{Style transfer generation property scores for a range of target properties sampled from EXP-cVAE-TF-Pol2 model.}
    \label{fig:TF-Range_Pol2}
\end{figure}

We repeated this experiment for the teacher forcing decoder (Figures \ref{fig:TF-Range_cVAE}, \ref{fig:TF-Range_KLD} and \ref{fig:TF-Range_Pol2}). The base model performs much better due to its increased focus on conditioning information. However we see that both regularizers still improve the range of achieved properties, diversity of solution modes and fidelity.

\newpage
\section{Dataset and Property Statistics}
\label{appndx:dataset}

\subsection{Dataset and Property Processing}

Training used properties and canonical SMILES extracted from the QM9 dataset file (provided as atom coordinates and types) using the RDKit chemo-informatics package (version 2023.09.6), ignoring molecules that resulted in invalid SMILES. The resulting SMILES were encoded as integers and augmented with start and end tokens. The remainder of each sequence was padded with 0 tokens.

\subsection{Data Splits}

We withheld 10000 random molecules from the dataset for the validation set (tracking progress and determining early stopping) and 10000 for the test set (performance metric computation). The remaining 113885 molecules were used as the training set. The same test-train-validation split was used to train all models, including the baseline model from \cite{Lim2018MolecularGM} and the LSTM model for the regularizers.

\subsection{Data Exploration}

\begin{figure}
    \centering
    \includegraphics[width=1.0\linewidth]{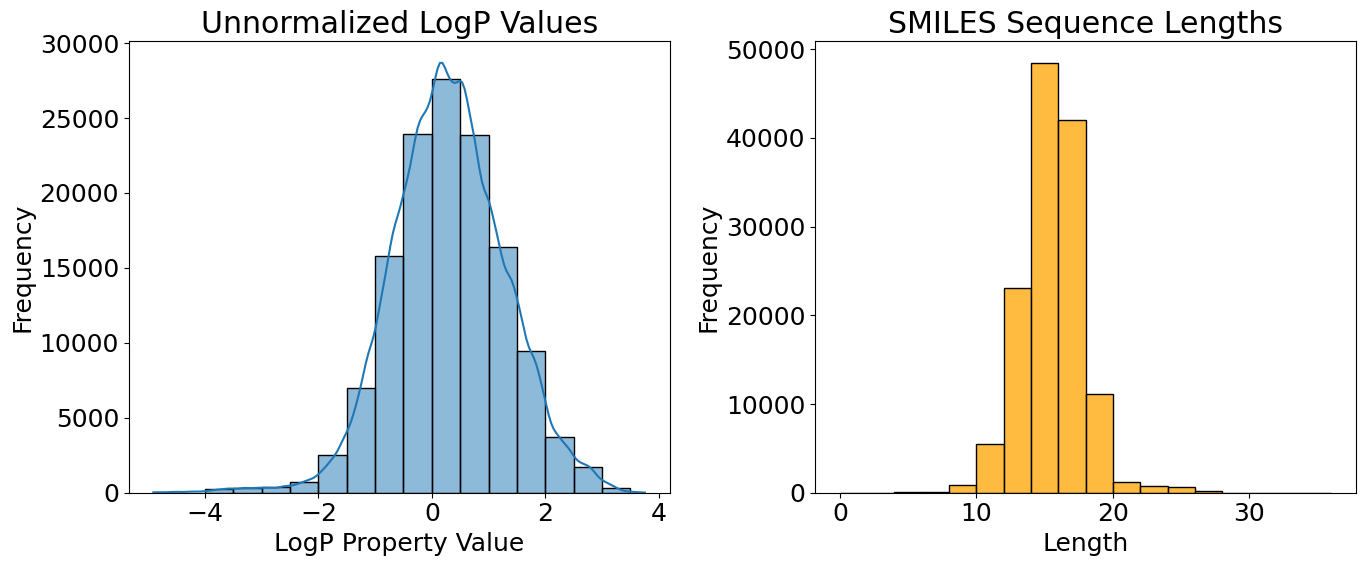}
    \caption{Summary statistics for QM9 dataset.}
    \label{fig:qm9_sumstats}
\end{figure}

All LogP values were normalized according to the distribution in the training-set (mean=0.2982, std=1.0009). The distributions of unnormalized LogP values and sequence lengths for the entire dataset are presented in Figure \ref{fig:qm9_sumstats}. The frequencies of various characters in the entire dataset can be found in Table \ref{character_statistics}.

\begin{table}[ht]
\caption{Symbol frequencies in entire QM9 dataset.}
\label{character_statistics}
\vskip 0.1in
\centering
\resizebox{\linewidth}{!}{%
\begin{tabular}{@{}c@{}c@{}c@{}}
\begin{minipage}[t]{0.33\linewidth}
\centering
\begin{tabular}{ccc}
\hline
\textbf{Symbol} & \textbf{Freq.} & \textbf{Norm. Freq.} \\
\hline
C & 767830 & 0.38 \\
1 & 259368 & 0.13 \\
O & 177823 & 0.09 \\
2 & 131751 & 0.07 \\
( & 118216 & 0.06 \\
) & 118216 & 0.06 \\
N & 98355 & 0.05 \\
\hline
\end{tabular}
\end{minipage}%
&
\begin{minipage}[t]{0.33\linewidth}
\centering
\begin{tabular}{ccc}
\hline
\textbf{Symbol} & \textbf{Freq.} & \textbf{Norm. Freq.} \\
\hline
= & 94597 & 0.05 \\
c & 78726 & 0.04 \\
n & 41409 & 0.02 \\
\# & 37027 & 0.02 \\
3 & 35634 & 0.02 \\
{[} & 12523 & 0.01 \\
{]} & 12523 & 0.01 \\
\hline
\end{tabular}
\end{minipage}%
&
\begin{minipage}[t]{0.33\linewidth}
\centering
\begin{tabular}{ccc}
\hline
\textbf{Symbol} & \textbf{Freq.} & \textbf{Norm. Freq.} \\
\hline
H & 10568 & 0.01 \\
o & 10174 & 0.01 \\
4 & 4952 & 0.00 \\
F & 3314 & 0.00 \\
- & 1974 & 0.00 \\
+ & 1847 & 0.00 \\
5 & 174 & 0.00 \\
\hline
\end{tabular}
\end{minipage}
\end{tabular}
}
\end{table}

\newpage
\section{Model Architecture and Training Parameters}
\label{appndx:parameters}

\subsection{Decoder Structure}
\label{appndx:decoder_structure}

Our experiments employ three different decoder structures: the one-shot decoder (cVAE), explicitly auto-regressive decoder without teacher forcing (EXP-cVAE) and explicit auto-regressive decoder with teacher-forcing:

\begin{itemize}
    \item \textbf{One-Shot}: The decoder does not accept the tokens generated from previous steps as input and propagates sequential information solely through the hidden state. Sequences are sampled from the model's predictions by independently sampling tokens with probabilities proportionate to each decoder output for each step in the sequence.  
    \item \textbf{Explicit/Auto-Regressive}: At each step in the sequence $t$, a token is sampled proportionately to the decoders outputs and provided as an input at $t+1$.
    \item \textbf{Explicit/AR with Teacher Forcing}: At each step in the sequence $t$, a token is sampled proportionately to the decoder outputs. The token provided as input to the decoder at $t+1$ is taken from the ground truth SMILES string. During inference the model is sampled in the same manner as the non-teacher forcing version (EXP-cVAE). 
    
\end{itemize}

\subsection{LSTM Parameters}

For the surrogate distribution $\tilde{p}(\mathbf{x}|\mathbf{y})$, we trained a conditional LSTM model with a maximum likelihood objective until convergence. The performance of this pre-trained model is shown in Table \ref{exp-gen}. The LSTM model consisted of an initial embedding layer (47x512) three stacked LSTM layers with input feature size 513 and output feature size 512 and a final linear layer with input size 512 and output size 47. Conditional information was provided by prepending it to the input at each time step. LSTM layers were trained with a dropout probability of 0.2. Initial and final layer weights were initialized using a xavier uniform initialization strategy. 

\subsection{VAE Architecture and Training}

All VAE models shared a common architecture and parameters (except for regularizer specific parameters and the latent variable dimensionality). Training was conducted for up to 500 epochs, with early stopping after 50 epochs without validation loss improvement. Regularizer weights were determined so that the initial regularizer loss component was approximately an order of magnitude smaller than the initial KL loss component. The ELBO KL loss component was annealed according to a sigmoid schedule (from 0.01 to 1.0) centered at epoch 29 and ending at epoch 120.

The VAE model encoder consisted of an initial embedding layer followed by three 1 dimensional convolutional layers of dimensions 47x9x9, 9x9x9 and 9x10x11, each followed by a ReLu activation function. These were fed into two linear layers, one for predicting the embedding mean (435x56) and one for the log variance (435x56). The state decoder consisted an initial embedding layer (47x56) followed by three stacked GRU layers (113x501, 501x501, 501x501) each followed and a final output layer (501x47). The models were trained with gradient clipping above set to magnitude 0.2 and a learning rate scheduler set to reduce learning rate on plateaus to a minimum of 1e-6. An overview of the training parameters is presented in Table \ref{tab:params}.

\begin{table}[ht]
    \caption{Training parameters for VAE models.}
    \label{tab:params}
    \vskip 0.1in
    \centering
    \begin{tabular}{cc}  
    \hline
        \textbf{Parameter} & \textbf{Value} \\
    \hline
         QM9 Latent Dim & 156 \\
         ZINC Latent Dim & 192 \\ 
         Optimizer & Adam \\
         Batch Size & 128 \\
         Learning Rate & 1e-4 \\
         ELBO KL & 0.001 to 1.0 \\
         Calib. Regularizer Weight & 0.1 \\
         Pol1 Regularizer Weight & 0.01\\
         Pol1 MC Sample Size for Latent Points & 4 \\
         Pol1 MC Sample Size for Decoder & 20 \\
         Pol2 Regularizer Weight & 0.01 \\
         Pol2/ MC Sample Size & 60 \\
    \hline
    \end{tabular}
\end{table}

\subsection{Implementation and Hardware Details}

The model was implemented using the PyTorch library (version 2.2.1) and trained on a NVIDIA GeForce RTX 4090 graphics card using CUDA version 12.3.

\newpage
\section{Metric Computation Details}
\label{appndx:metric_computation}

Reconstruction was computed from by variationally encoding a 1000 test set SMILES 5 times each. These were then decoder a further 5 times. The reported reconstruction score is the proportion of resulting SMILES that are valid and who's canonical form is identical to the input SMILES."

\subsection{Conditional Generation Metrics}

Conditional generation metrics were computed by sampling 1000 latent points $\bz$ and equipping each with a random $\by$ from the test set. Each pair was decoded 5 separate times to produce the final sample. We evaluated conditional generation with the following metrics:

\begin{itemize}
    \item \textbf{Validity}: The proportion of valid molecules in the sample. We considered SMILES successfully parsed by RdKit to be valid.
    \item \textbf{Uniqueness}: The number of unique valid molecules among all valid molecules. Uniqueness was computed from valid SMILES canonicalized by RdKit.
    \item \textbf{Novelty}: The proportion of canonical forms of valid SMILES not present in the training dataset.
    \item \textbf{Property MAE}: The mean absolute error between conditioning properties and properties computed from the parsed valid SMILES. 
\end{itemize}

\subsection{Style Transfer Metrics}

"Style transfer (molecular optimization) metrics were computed by variationally encoding 2500 test set SMILES (the source SMILES). Each of these was then equipped with an independently sampled test set property (the target property), and decoded to produce the final generated molecule. The following metrics were used to evaluate style transfer performance:

\begin{itemize}
    \item \textbf{PCT Valid: } The proportion of generated molecules that are successfully parsed by RdKit.
    \item \textbf{PCT Fail: } The proportion of generated molecules that are identical to the source molecules. In these cases the model has failed to consider the conditional information and change the molecule.
    \item \textbf{MAE:} The MAE between the target property and property score computed from valid generated SMILES.
    \item \textbf{Tan:} The structural similarity between the source and generated molecule. Computed as the Tanimoto similarity between the Morgan fingerprints (with radius 3) of each molecule and ranging from 0 to 1.
\end{itemize}

\subsection{Baselines for Tanimoto Similarity}
\label{appdnx:tanimoto}

The first Tanimoto baseline was computed as the average similarity between the set SMILES generated through style transfer and a paired set of generations using random latent points instead of the encoded source molecules. The second Tanimoto baseline compared the similarity of two sets of style transferred SMILES with a shared set of target properties.

\subsection{Baseline for Property MAE}
We determined a baseline upper-bound for the MAE between two molecules from the QM9 setting at $1.1213$. It was computed as the MAE between each possible pair of test set molecules and represents a case where property information has been completely ignored.

\subsection{Comparative Evaluation of Models}
Both conditional generation and molecular optimization are multifaceted problems. For example, if conditional generation model fails to generate either valid, unique or novel molecules, it would not be practical for use in drug design. Similarly, molecular optimization involves a trade off between structural similarity to the source and respecting the conditioning information. This trade off necessarily exists because a molecules property and structure cannot be completely disentangled and consequently it is impossible to change a molecules property without affecting its structure. The degree to which this is possible also depends on both the specific molecule under consideration and the nature of the property being considered. As a result it is difficult to give a general ranking between models where one is not Pareto dominant, and we must instead consider the trade off between the two.


\end{document}